\documentclass[review,12pt]{elsarticle}

\usepackage{amssymb}
\usepackage{hyperref}
\usepackage{amsmath}
\usepackage[margin=2cm]{geometry}% TO PLAY WITH THE MARGINS
\usepackage{subcaption}
\usepackage{float}
\usepackage{color}
\biboptions{sort&compress}
\journal{Advances in Engineering Software}
\begin{document}
\begin{frontmatter}
\title{UMAT4COMSOL: An Abaqus user material (UMAT) subroutine wrapper for COMSOL}
%\title{UMAT4COMSOL: An Abaqus \textit{UMAT} wrapper to COMSOL, applications to elastoplasticity, hyperelasticity and crystal plasticity}
\author[add1]{Sergio Lucarini\corref{corr1}}
\ead{s.lucarini@imperial.ac.uk}
\author[Ox,add1]{Emilio Mart{\' i}nez-Pa{\~ n}eda\corref{corr1}}
\ead{emilio.martinez-paneda@eng.ox.ac.uk}

\address[add1]{Department of Civil and Environmental Engineering, Imperial College, London SW7 2AZ, UK}
\address[Ox]{Department of Engineering Science, University of Oxford, Oxford OX1 3PJ, UK}
\cortext[corr1]{Corresponding authors}

\begin{abstract}
We present a wrapper that allows Abaqus user material subroutines (\textit{UMAT}s) to be used as an \textit{External Material} library in the software COMSOL Multiphysics. The wrapper, written in C language, transforms COMSOL's external material subroutine inputs and outputs into Fortran-coded Abaqus \textit{UMAT} inputs and outputs, by means of a consistent variable transformation. This significantly facilitates conducting coupled, multi-physics studies employing the advanced material models that the solid mechanics community has developed over the past decades. We exemplify the potential of our new framework, UMAT4COMSOL, by conducting numerical experiments in the areas of elastoplasticity, hyperelasticity and crystal plasticity. The source code, detailed documentation and example tutorials are made freely available to download at \url{www.empaneda.com/codes}.\\
\end{abstract}
\end{frontmatter}

\section{Introduction}
Coupling multiple physical phenomena in continuum solids is a major research focus \cite{RILEM2021,Mianroodi2022}. The mechanical behaviour of structures and industrial components is relatively well understood in inert environments but challenges arise as a result of material-environment interactions. Multi-physics structural integrity problems such as hydrogen embrittlement \cite{AM2016,Gobbi2019}, corrosion \cite{JMPS2021,Ansari2021}, and oxidation-assisted fatigue \cite{Reuchet1983,LeoPrakash2009} continue to challenge scientists and engineers. Predicting these structural integrity problems requires coupling state-of-the-art material models with equations describing chemical, electrical and thermal phenomena \cite{CMAME2018,JMPS2020,Valverde-Gonzalez2022,Cui2022,hageman2023stabilising}. In addition, numerous modern devices and concepts involve multi-physics environments and mechanical loads, which are either applied externally or induced by other physical phenomena, such as thermal expansion, chemical strains or magnetic forces. Examples include Li-Ion batteries \cite{Boyce2022,Ai2022}, hydrogels \cite{Pan2022}, magnetorheological elastomers \cite{Rambausek2022,Moreno-Mateos2023}, and piezo-electric and piezo-resistive materials \cite{Wu2021b,Quinteros2023}, to name a few. As a result, there is a pressing need to develop computational tools for characterising the coupled behaviour of materials in multiphysics environments \cite{Nie2010,Patzak2013,Zhao2019}.\\
%Add JMPS2022b later and Chuanjie's COMSOL paper

Abaqus \cite{Abaqus2022} and COMSOL \cite{Comsol2022} have arguably been the most popular finite element packages for materials modelling and multi-physics simulations, respectively. For decades, the solid mechanics community has developed Abaqus user material subroutines (\textit{UMAT}s) to numerically implement new and advanced material models; e.g., in the context of hyperelasticity \citep{Sun2008SL}, elastoplasticity \citep{Ramasubramanian2007}, damage \cite{Materials2021}, and crystal plasticity \cite{Huang1991,Kysar1997}. Abaqus \textit{UMAT} subroutines are Fortran codes that describe constitutive material behaviour under either the small strains or finite strains convention. On the other hand, COMSOL stands out for its ability to handle coupled systems of partial differential equations, with multiple built-in physics modules that encompass most physical phenomena across the disciplines of chemistry, fluid flow, heat transfer, electromagnetism, structural mechanics and acoustics. Given the growing interest in material problems in multi-physics environments, there is great benefit in coupling these two tools. However, the COMSOL materials library does not currently support the automatic use of Abaqus \textit{UMAT} subroutines as material models. Available materials in COMSOL are either built-in or can be programmed in C language as a user-defined \textit{External Material} library. Therefore, our aim is to develop and share a wrapper that can take advantage of pre-programmed Abaqus \textit{UMAT} material models and enable their use as \textit{External Material} libraries in COMSOL. Both small and finite strains are considered, providing a general framework that can leverage the notable efforts made in materials model development with Abaqus \textit{UMAT}s, while also taking advantage of the versatility of COMSOL for multi-physics simulations. In this way, the present work contributes to ongoing efforts in the community aimed at connecting computational platforms \cite{segurado2013simulation,helfer2015introducing,portillo2017muesli,AES2017,Helfer2020}.\\

The remainder of this paper is organised as follows. First, in Section \ref{Sec:SoftwareDescription}, we proceed to describe the characteristics of the software that we have developed, UMAT4COMSOL. Then, in Section \ref{Sec:Usage}, usage instructions are provided. The potential of UMAT4COMSOL is demonstrated in Section \ref{Sec:Results} through examples encompassing small strain elastoplasticity, large strain hyperelasticity and crystal plasticity. Finally, concluding remarks end the manuscript in Section \ref{Sec:Conclusions}.

\section{Software description: UMAT4COMSOL}
\label{Sec:SoftwareDescription}

UMAT4COMSOL is a C-coded subroutine that functions as a user-defined \textit{External Material} in COMSOL, so as to describe the local mechanical behaviour of solid domains. The subroutine contains a call to an external Fortran-coded subroutine that corresponds to the Abaqus \textit{UMAT}. During this process, the local state is inputted, and the constitutive equations describing the material behaviour are computed, with their outputs being returned to COMSOL for solving the multiphysics global problem.\\

UMAT4COMSOL's primary internal features are described below, starting with a concise explanation of the input requirements for both COMSOL \textit{External Material}s and Abaqus \textit{UMAT}s. Subsequently, the process of transferring input parameters from COMSOL \textit{External Materials} to Abaqus \textit{UMAT}s is elaborated upon. Finally, the output procedure, which involves a more detailed understanding of the various mathematical methods, is described.

\subsection{Material models inputs and outputs}

When modelling the mechanical behaviour of materials, the constitutive equations characterise the relation between the measure of the deformation and the stress state of the material. These equations can be linear, non-linear, time-dependent, and history-dependent.\\

The input to the material constitutive equations is typically a deformation measure. In the small strain formalism, the strain tensor $\varepsilon_{ij}$ is used as the deformation measure, and is defined in terms of the displacement vector $u_i$ as
\begin{equation}
    \varepsilon_{ij}=\frac{1}{2}\left( u_{i,j} + u_{j,i} \right) \, ,
\end{equation}
while the deformation gradient $F_{ij}$, given by
\begin{equation}\label{eq:defgra}
    F_{ij}= \delta_{ij} + u_{i,j} \, ,
\end{equation}
is typically adopted when considering finite strains. Here, $\delta_{ij}$ denotes the Dirac delta function and the derivative definition in Eq. \eqref{eq:defgra} is defined in the reference configuration. Another class of inputs are the material properties, which are typically constant values. Additionally, the constitutive equations can be posed as a function of other strain measures dependent on the aforementioned ones, such as the left Cauchy-Green or Green-Lagrange tensors. In the case of history-dependent material behaviour, one should also consider a state variable vector $\alpha_i$, that stores history information of the state of the material. Finally, in time-dependent problems, the time increment value is also stored, so as to calculate the strain rate measures and the evolution of state variables.\\

The outcome of the constitutive calculations is a calculated stress measure, which will be translated to forces at the element level. In the case of non-linear analyses, the mechanical equilibrium equation is usually linearised, requiring the definition of a consistent tangent modulus, which corresponds to the derivative of the adopted stress measure with respect to the adopted strain measure.

\subsection{COMSOL External Material}

COMSOL \textit{External Material}s are C-coded subroutines that are calculated at the Gauss point level on each iteration using a subroutine called \texttt{eval}. The inputs include the strain/deformation gradient trials \verb|E|/\verb|Fl|, along with the properties vector \verb|par| and the deformation gradient at the previous time increment \verb|FlOld|. The (pseudo)time steps/increments are defined in the \verb|delta| variable, and a set of state variables are passed as a vector \verb|state|. Once the constitutive equations are calculated in the core code, the results obtained for each (pseudo)time step/increment are passed using the stress vector \verb|S|, the consistent tangent \verb|D|/\verb|Jac|, and the evolved state variables \verb|state|.

\subsection{Abaqus user material (UMAT) subroutine}

Abaqus \textit{UMAT}s are Fortran-coded subroutines that calculate the constitutive equations at the Gauss point level on every iteration. The inputs for the mechanical problem include the trial incremental quantities of strain/deformation trials at the current increment (\verb|DSTRAN|/\verb|DFGRD1|), the total strain/deformation measures at the end of the previous increment (\verb|STRAN|/\verb|DFGRD0|), the stresses at the previous time increment \verb|STRESS|, and the material properties, which are contained in the \verb|PROPS| vector. The (pseudo)time increments are defined in the \verb|DTIME| variable, together with the current (pseudo)time value (\verb|TIME|), and a set of state variables is passed as a vector in \verb|STATEV|. Also, some constitutive relations in a finite strain context need the incremental rotation tensor variable \verb|DROT|. After calculating the constitutive equations, the key outputs at the end of the (pseudo)time increment are the updated stress vector \verb|STRESS|, the consistent tangent matrix \verb|DDSDDE|, and the updated state variables vector \verb|STATEV|.

\subsection{Transfer of inputs}
\label{sec:inp}

The code initially transforms COMSOL's input variables to conform to the convention of Abaqus \textit{UMAT} subroutines, which requires a change in the strain format. Specifically, two aspects must be taken care of in small strain problems. One is that Abaqus stores shear strains as engineering shear strains; $\gamma_{xy}=2\varepsilon_{xy}$. A second one is that the ordering of the shear components is different. Accordingly, one must change the COMSOL input strain tensor  (\verb|E| variable) to fit the \textit{UMAT} convention, such that
\begin{equation}\label{eq:voi}
\varepsilon_{xx},\varepsilon_{yy},\varepsilon_{zz},\varepsilon_{yz},\varepsilon_{xz},\varepsilon_{xy} \rightarrow \varepsilon_{xx},\varepsilon_{yy},\varepsilon_{zz},\gamma_{xy},\gamma_{xz},\gamma_{yz}
\end{equation}
where $\gamma_{\square}=2\varepsilon_{\square}$ and the left-hand side corresponds to the COMSOL notation. Here, one should note that COMSOL's \verb|E| vector contains the total trial strains. Consequently, the incremental trial strains vector, \verb|DSTRAN|, is computed by subtracting from \verb|E| the total strain at the end of the previous increment, which has been stored as state variables and corresponds to the variable \verb|STRAN| in the \textit{UMAT} convention. Also, the stress at the previous time increment is saved in the state variable vector and assigned to the \textit{UMAT} variable \verb|STRESS| before computing the constitutive equations. For the finite strains version, the \textit{UMAT} requires the deformation gradient $F_{ij}$ before and after the (pseudo)time increment. COMSOL provides the deformation gradients \verb|FlOld| and \verb|Fl| to the external material subroutine as a C-order matrix. These matrices are transformed into a Fortran-order matrix and passed to the \textit{UMAT} as \verb|DFGRD0| and \verb|DFGRD1|. Additionally, the incremental strain \verb|DSTRAN| ($\Delta \varepsilon_{ij}$) and incremental rotation \verb|DROT| ($R^{\Delta t}_{ij}$) are defined as auxiliary input variables. The \verb|DSTRAN|  components are computed via 
\begin{equation}
      \Delta \varepsilon_{ij}=1/2\left( \Delta F_{ip}  F_{t+\Delta t \ | \ pj}^{-1} + F_{t+\Delta t \ | \ pi}^{-1} \Delta F_{jp}\right) \, , 
\end{equation}

\noindent defining $\Delta F_{ij}=F_{t+\Delta t \ | \ ij}-F_{t \ | \ ij}$ and forming a vector by using Eq. (\ref{eq:voi}). The variable \verb|DROT| is defined by 
\begin{equation}
 F^{\Delta t}_{ij}=R^{\Delta t}_{ip} U^{\Delta t}_{pj},   
\end{equation}
standing for the polar decomposition of $F^{\Delta t}_{ij}=F_{t+\Delta t \ | \ ip}F_{t \ | \ pj}^{-1}$. \color{black}Finally, the values of the current (pseudo)time and (pseudo)time increment are obtained by making use of COMSOL's \verb|delta| variable and updating a state variable. The COMSOL variable \verb|delta| contains the current time increment and thus corresponds to the \textit{UMAT} variable \verb|DTIME|. Since the \textit{UMAT} subroutine also takes as input the total time at the beginning of the increment (\verb|TIME| variable), this is stored as the first component of the \verb|STATEV| vector, which is updated on every increment based on \verb|delta|. 

\subsection{Transfer of outputs}
\label{sec:out}

The stress vector and the consistent tangent matrix are the outputs of the \textit{UMAT}s. In the small strain version, the stress variable (\verb|STRESS|) can be directly transformed by rearranging stress vector shear components, as in Eq. (\ref{eq:voi}), to match the required output format for the COMSOL solver (\verb|S| variable). In a similar fashion, the consistent tangent matrix \verb|DDSDDE| is reordered to match the arrangement of the shear components, as per (Eq. \ref{eq:voi}), and then transposed to convert from a Fortran-ordered matrix to a C-ordered matrix \verb|D|.\\

For the finite strain case, the outputs of the \textit{UMAT} subroutine (the stresses and the consistent tangent matrix) need to be transformed before being transferred. In regards to the stresses, Abaqus \textit{UMAT} outputs the Cauchy stress tensor $\sigma_{ij}$ in the variable \verb|STRESS| (as a vector, using Voigt notation), while COMSOL works with the second Piola-Kirchhoff stress $S_{ij}$, using the variable \verb|S|. The transformation, \verb|STRESS| $\rightarrow$ \verb|S|, is defined using the following relation
\begin{equation}\label{eq:2ndpk}
S_{ij}= J F^{-1}_{iq} \sigma_{qp} F^{-T}_{jp}\text{ ,}
\end{equation}
where $J=\det\left(F_{ij}\right)$. \\

As described below, the transformation of the consistent tangent tensor (\verb|DDSDDE| $\rightarrow$ \verb|Jac|) in a finite strains context requires careful consideration. This is also a key step to guarantee the appropriate convergence rate. The material Jacobian obtained from the Abaqus \textit{UMAT} user subroutine at finite strains is the tangent modulus tensor for the Jaumann rate of the Kirchhoff stress, denoted by $C^{Abaqus}_{ijkl}$, which is a fourth-order tensor defined as
\begin{equation}\label{eq:abatang}
\overset{\nabla}{\tau}_{ij}= \dot{\tau}_{ij} - w_{ip} \tau_{pj} - \tau_{ip} w_{jp} = J \ C^{\text{Abaqus}}_{ijkl}  d_{kl}
\end{equation}
where $\tau_{ij}$ denotes the Kirchhoff stress, $\overset{\nabla}{}$ is the Jaumann rate, and $w_{ij}$ and $d_{ij}$ are the spin tensor and stretch tensor, respectively. In contrast, the COMSOL framework requires the material tangent, $K^{\text{COMSOL}}_{ijkl}$, which is given by
\begin{equation}\label{eq:comtang}
\dot{S}_{ij} = K^{\text{COMSOL}}_{ijkl} \dot{F}_{kl}\text{ .}
\end{equation}
We shall then derive an explicit expression that relates the two fourth-order tensors, $C^{\text{Abaqus}}_{ijkl}$ and $K^{\text{COMSOL}}_{ijkl}$. Combining Eqs. \eqref{eq:2ndpk} and \eqref{eq:comtang}, one reaches
\begin{equation}\label{eq:proddivision}
K^{\text{COMSOL}}_{ijkl}=\frac{\partial S_{ij}}{\partial F_{kl}}=
\frac{\partial \left( J F^{-1}_{iq} \sigma_{qp} F^{-1}_{jp}\right)}{\partial F_{kl}}=
\frac{\partial \left( F^{-1}_{iq} \tau_{qp} F^{-1}_{jp}\right)}{\partial F_{kl}}.
\end{equation}
Expanding the derivative of the product in Eq. \eqref{eq:proddivision} and expressing it in index notation, we obtain:
\begin{equation}\label{eq:indexx}
K_{ijkl}^{\text{COMSOL}}= \frac{\partial S_{ij}}{\partial F_{kl}}=
\frac{\partial F^{-1}_{iq}}{\partial F_{kl}}\tau_{qp}F^{-1}_{jp}+
F^{-1}_{iq}\frac{\partial \tau_{qp}}{\partial F_{kl}} F^{-1}_{jp}+
F^{-1}_{iq}\tau_{qp}\frac{\partial F^{-1}_{jp}}{\partial F_{kl}}
\end{equation}
This equation expresses the components of the COMSOL material tangent, $K^{\text{COMSOL}}_{ijkl}$, as a function of components that can be obtained using the definition of the Abaqus tangent modulus tensor $C^{\text{Abaqus}}_{ijkl}$, see Eq. \eqref{eq:abatang}. The first and third terms in Eq. \eqref{eq:indexx} are the derivatives of the inverse of a tensor with respect to itself: $\partial F^{-1}_{jp}/\partial F_{kl}=-F^{-1}_{lp}F^{-1}_{jk}$. To obtain the second term in Eq. \eqref{eq:indexx}, we linearise the rate quantities from Eq. \eqref{eq:abatang} by multiplying by an infinitesimal time increment and reordering terms, yielding
\begin{equation}\label{eq:dtau}
\delta\tau_{ij}  = J \ C^{\text{Abaqus}}_{ijkl}  \delta d_{kl} + \delta w_{ip} \tau_{pj} + \tau_{ip} \delta w_{jp}\text{ ,}
\end{equation}
where $\delta d_{ij}$ and $\delta w_{ij}$ are obtained as functions of $F_{ij}$ and $\delta F_{ij}$ via
\begin{equation}\label{eq:basic}
\delta d_{ij}=\frac{1}{2} \delta F_{ip} F^{-1}_{pj} + \frac{1}{2} \delta F_{jp} F^{-1}_{pi} 
\text{ ,  }
\delta w_{ij}=\frac{1}{2} \delta F_{ip} F^{-1}_{pj} - \frac{1}{2} \delta F_{jp} F^{-1}_{pi} \text{ .}
\end{equation}

The Kirchhoff stress is also linearised with respect to the perturbation of the deformation gradient as
\begin{equation}\label{eq:lintau}
\delta \tau_{ip} =
\frac{\partial \tau_{ip}}{\partial F_{kl}} \delta F_{kl}.
\end{equation}
where $\delta\tau_{ij}$ and $\delta F_{ij}$ are the Kirchhoff stress and the deformation gradient perturbations, respectively. Inserting the expressions of Eq. \eqref{eq:basic} into Eq. \eqref{eq:dtau}, we obtain 
\begin{equation}\label{eq:indexini}
\begin{aligned}
\delta\tau_{ip} = \frac{J}{2} C^{\text{Abaqus}}_{ipkm} \delta F _{kl} F^{-1}_{lm} + \frac{J}{2} C^{\text{Abaqus}}_{ipmk}  F^{-1}_{lm} \delta F _{kl}  
\\ 
+\frac{1}{2} I_{ipkq} \delta F_{kl} F^{-1}_{lm} \tau_{mq} -
\frac{1}{2} I_{ipmq} F^{-1}_{lm} \delta F_{kl} \tau_{kq} 
\\
+\frac{1}{2} I_{ipqk} \tau_{qm} F^{-1}_{lm} \delta F_{kl} -
\frac{1}{2} I_{ipqm} \tau_{qk} F^{-1}_{lm} \delta F_{kl}
\end{aligned}
\end{equation}
which we can be used to obtain $\partial \tau_{ip} / \partial F_{kl}$, considering Eq. \eqref{eq:lintau}. By considering the minor symmetries of $C^{\text{Abaqus}}_{ijkl}$ and the symmetry of $\tau_{ij}$, we can simplify the expression and arrive at the following definition of $\partial \tau_{ip} / \partial F_{kl}$
\begin{equation}\label{eq:tau_f_function_Cab}
\frac{\partial \tau_{ip}}{\partial F_{kl}} =
J \ C^{\text{Abaqus}}_{ipkm} F^{-1}_{lm} +
\frac{1}{2} \delta_{ik} F^{-1}_{lm} \tau_{mp} -
\frac{1}{2} F^{-1}_{li} \tau_{kp} +
\frac{1}{2} \delta_{pk} F^{-1}_{lm} \tau_{im} -
\frac{1}{2} F^{-1}_{lp} \tau_{ik}\text{ .}
\end{equation}

The relation between the COMSOL and Abaqus material tangents is then obtained by inserting Eq. (\ref{eq:tau_f_function_Cab}) into Eq. (\ref{eq:indexx}), rendering
\begin{equation}
\begin{aligned}
K_{ijkl}^{\text{COMSOL}}= &
- F^{-1}_{lq} F^{-1}_{ik} \tau_{qp} F^{-1}_{jp} + F^{-1}_{iq} \Bigl( J \ C^{\text{Abaqus}}_{qpkm} F^{-1}_{lm} +
\frac{1}{2} \delta_{qk} F^{-1}_{lm} \tau_{mp} \\
& - \frac{1}{2} F^{-1}_{lq} \tau_{kp} +
\frac{1}{2} \delta_{pk} F^{-1}_{lm} \tau_{qm}-\frac{1}{2} F^{-1}_{lp} \tau_{qk} \Bigr) F^{-1}_{jp}-
F^{-1}_{iq} \tau_{qp} F^{-1}_{lp} F^{-1}_{jk}\text{ .}
\end{aligned}
\end{equation}
The resulting COMSOL consistent tangent can then be transformed to Voigt notation and transposed from Fortran-order to C-order matrix, which is the required output to be used by the COMSOL solver.

\section{Usage instructions}
\label{Sec:Usage}

This section provides a brief description of the source code assembly, which aims to aid in understanding the software and enable personalised development. The relevant operations to use the wrapper are also described. UMAT4COMSOL is made freely available to download at \url{https://github.com/sergiolucarini}, where future forks will be allowed to enable community developments, and \url{www.empaneda.com/codes}. 

\subsection{Organization of the source code}

UMAT4COMSOL is a C code that defines the \texttt{eval} function required in COMSOL for user-defined \textit{External material}s. Two versions are available: a small strain version and a finite strain version. Both versions have a similar structure. One of the key elements of the code is an external \texttt{umat} function, which is linked to the Fortran \textit{UMAT} subroutine. The first part of the code handles this \texttt{umat} function, which is followed by the standard header for a COMSOL user-defined external material subroutine, as per the COMSOL reference manual. Then, input variables are transformed using the operations described in Section \ref{sec:inp}. The core code calls the Abaqus \textit{UMAT} subroutine to compute the constitutive equations, and the last part of the code converts the output data, as explained in Section \ref{sec:out}, which is then passed to the COMSOL solver.

\subsection{Main usage}

The main file is a wrapper in C language called \texttt{UMAT4COMSOL.c}. Once compiled, it can be used as a user-defined \textit{External Material} library in COMSOL. For the compilation process, the \texttt{UMAT4COMSOL.c} and \texttt{umat.f} files should be in the same folder. The following commands can be used to compile the code:
\begin{verbatim}
gfortran -c umat.f
gcc -c UMAT4COMSOL.c
gcc -shared -o extmat.dll umat.o UMAT4COMSOL.o -lgfortran -lquadmath
\end{verbatim}

\noindent Here, the compilation process uses open-source compilers, but other Fortran and C compilers can also be used as long as the equivalent compiling flags are used. Note that the scheme can accommodate \textit{UMAT} subroutines written in both fixed Fortran (FORTRAN 77) and free Fortran (Fortran 90). The above commands generate two objects and a dynamic library called \texttt{extmat.dll} which is required for running simulations in COMSOL. If Linux is used, the extension of the library should be ``.so" (i.e., \texttt{extmat.so}). It should also be noted that, for finite strain problems, a different wrapper is provided, named \texttt{UMAT4COMSOLfinite.c}.\\

To run non-linear simulations with external materials in COMSOL, one can use either the \emph{Time-Dependent Solver} or the \textit{Stationary Solver}. For small strains, the material should be introduced as a \emph{General stress-strain} relation, and for finite strains, as a \emph{General stress-deformation} relation. Then, the properties vector needs to be included in the Material model parameters vector, and the number of state variables and the initialisation values of the vector must be given. Finally, the path to the compiled dynamic library (\texttt{extmat.dll}) must be entered in the \textit{External Material Library}'s field and imported into the COMSOL model.\\

To use external materials in the \textit{Solid Mechanics} module, an \textit{External Stress-Strain} relation node must be created for each phase in the domain and linked to the corresponding \textit{External Material} created. After setting the boundary conditions and solver settings, the simulation is ready to be run and post-processed.

\subsection{Sofware aspects}

\begin{figure}[H]\centering
\includegraphics[width=0.94\textwidth]{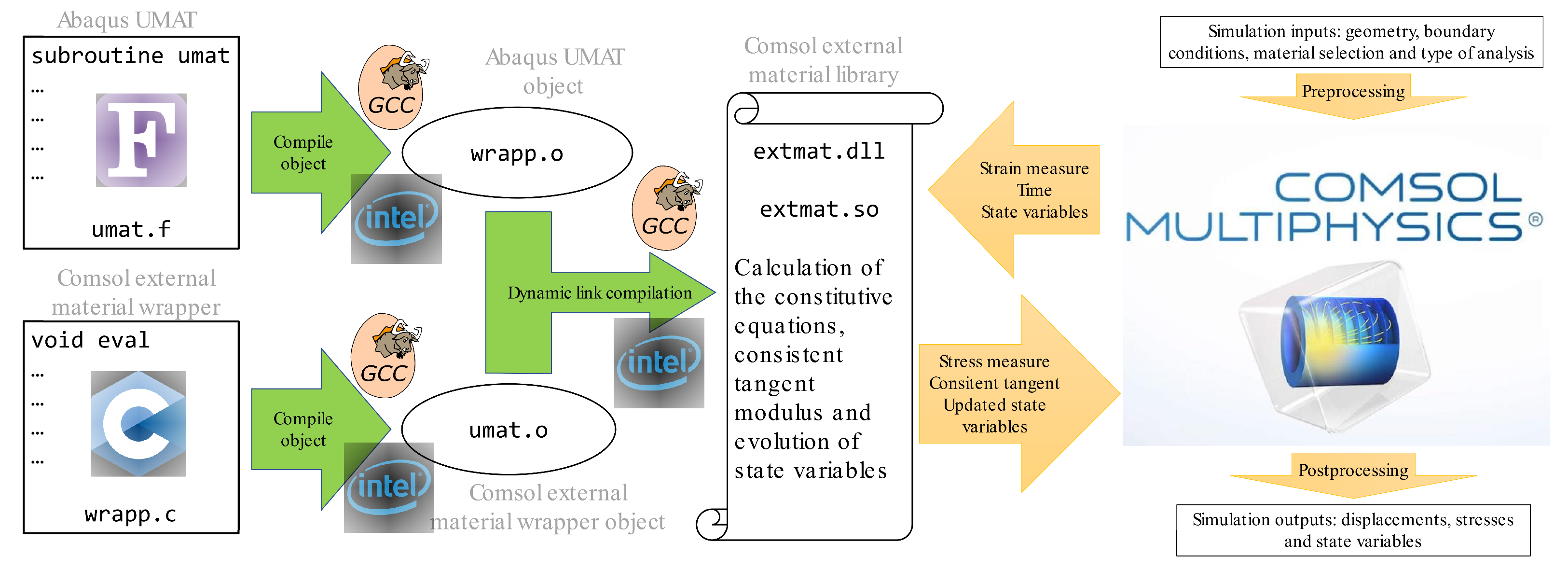}
\caption{Flow chart of UMAT4COMSOL}\label{fig:flow}
\end{figure}

As stated previously, this wrapper relies on the definition of a COMSOL \textit{External Material} library. A comprehensive scheme of the functioning of the framework is shown in Fig. \ref{fig:flow}. The procedure imports a compiled object, which encompasses Fortran and C code. The compiler choice is up to the user, taking into account the particularities of each compiler.\\

Once the compiled library is associated with the COMSOL input file, the wrapper will act as any material in COMSOL, enabling the use of the advanced functionalities that COMSOL offers. For non-linear analyses, COMSOL will access the \textit{External Material} library at least twice in each iteration, once for obtaining the stress and a second time for obtaining the consistent tangent. This makes the COMSOL analysis less efficient compared to Abaqus since the latter is optimised to extract both at once.

\section{Representative results and applications}
\label{Sec:Results}

To demonstrate the capabilities of UMAT4COMSOL, we provide representative results in the context of three areas of particular interest: elastoplasticity, large strain hyperelasticity and crystal plasticity; validating the outputs against Abaqus calculations. Thus, we first predict the behaviour of an elastoplastic plane stress holed plate undergoing small strains (Section \ref{Sec:Elastoplasticity}). Then, in Section \ref{Sec:Hyperelastic}, we verify the finite strains implementation by considering a neo-Hookean hyperelastic model. Finally, in Section \ref{Sec:CrystalPlasticity}, we employ an advanced crystal plasticity model to showcase UMAT4COMSOL potential in handling complex constitutive equations. These are just three examples of the capabilities of UMAT4COMSOL but its applicability is universal, as it provides a non-intrusive connection between COMSOL and any material models available as Abaqus \textit{UMAT}s. As an example of its potential, a fourth case study is included where a coupled problem is solved - the transport of hydrogen in a single crystal.\\ 

Of interest to all of these studies is the description of the tolerance criteria employed in Abaqus and COMSOL, as well as their comparison. Consider first the case of Abaqus. The tolerances for non-linear analysis in Abaqus are expressed in terms of maximum residual forces, such that equilibrium is achieved when this value is below a tolerance $\mathrm{tol}^{\text{Abaqus}}$. The inequality reads as
\begin{equation}\label{eq:abatol}
\mathrm{tol}^{\text{Abaqus}} > \frac{r^{max}}{\widetilde{q}} \text{ ,}
\end{equation}
where $r^{max}$ is the maximum value of the residual force vector and $\widetilde{q}$ is the overall time-averaged value of the 
spatially averaged force over the entire model. On the other hand, COMSOL considers that convergence has been achieved when the relative tolerance $\mathrm{tol}^{\text{COMSOL}}$ exceeds the relative error computed as the weighted Euclidean norm. That is,
\begin{equation}\label{eq:comtol}
\mathrm{tol}^{\text{COMSOL}} > \sqrt{\frac{1}{N}\sum_{i=1}^N \left(\frac{|E_i|}{\widetilde{W}}\right)^2} \text{ ,}
\end{equation}
where $E_i$ is the estimated solution error vector of the $i$-th degree of freedom, $N$ is the number of degrees of freedom and $\widetilde{W}$ is the weight, which is determined by $\widetilde{W} = \max \left( |U_i|, S \right)$, where $|U_i|$ is the absolute value of the solution at that specific degree of freedom and $S$ the absolute value of the average of the solution vector. For the numerical benchmarks a non-linear tolerance of $\mathrm{tol}^{\text{Abaqus}}=5 \times 10^{-3}$ and $\mathrm{tol}^{\text{COMSOL}}=1 \times 10^{-3}$ has been set for the Abaqus and COMSOL solvers, respectively. These are the default tolerances for both solvers. It is worth noting that while Abaqus also has a solution-based tolerance criterion, the residual-based one given in Eq. (\ref{eq:abatol}) is typically more restrictive.

\subsection{Small strain elastoplastic model}
\label{Sec:Elastoplasticity}

\begin{figure}[ht!]\centering
\includegraphics[width=1\textwidth]{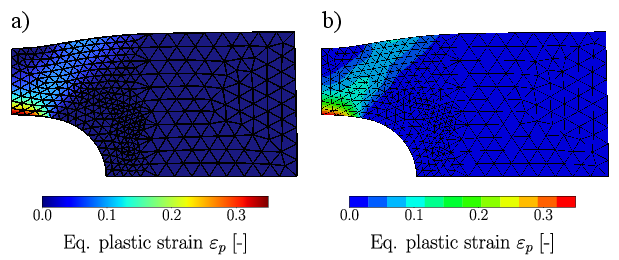}
\caption{Mechanical response of an elastoplastic holed plate: contour plots of the equivalent plastic strain. Results obtained: (a) with Abaqus, using an elastoplastic \emph{UMAT}, and (b) with COMSOL, using the same \emph{UMAT} and UMAT4COMSOL. The deformation has been scaled by a factor of 20.}
\label{fig:evpfield}
\end{figure}

\begin{figure}[H]\centering
\includegraphics[width=0.95\textwidth]{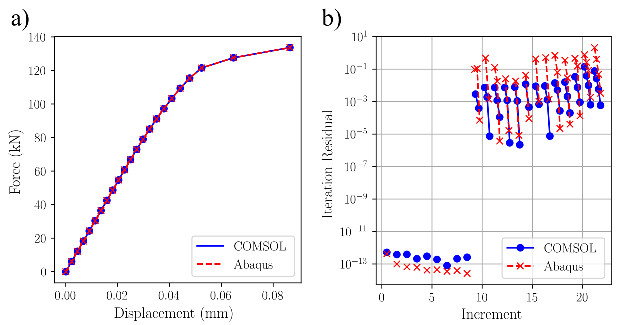}
\caption{Mechanical response of an elastoplastic holed plate: (a) Force versus displacement response, predicted with Abaqus (+\emph{UMAT}) and with COMSOL (+\emph{UMAT} and UMAT4COMSOL); and (b) convergence plots for both COMSOL and Abaqus solvers, showing the magnitude of the residual for each iteration as a function of the load increment. In (b), iterations are represented with equispaced subdivisions within each increment interval.}
\label{fig:evpssc}
\end{figure}

In this example, we simulate the mechanical behaviour of a rectangular plate with a circular hole that is undergoing plastic deformation. The geometry and loading conditions are similar to the benchmark problem presented in Ref. \cite{Zienkiewicz2013}. The plate has a width of 36 mm and a height of 20 mm, and contains a hole of 5 mm radius at its centre. Due to symmetry, only one quarter of the model is simulated, with symmetry boundary conditions being applied on the left and bottom edges. The thickness of the plate is much smaller than the other dimensions, and the loads are confined to the plate plane, so plane stress conditions are assumed. The right edge of the plate is subjected to a traction that ramps linearly up to a maximum of $133.65$ MPa. The plate is made of an isotropic elastoplastic material whose elastic response is characterised by a value of Young's modulus of $E = 70$ GPa and a Poisson's ratio of $\nu = 0.2$. In all case studies, we select the type of finite element consistently across platforms. In this specific benchmark, first-order plane stress Lagrange triangular elements with reduced integration are used in COMSOL, and accordingly, so-called CPS6 elements are used in the Abaqus calculations - the analogous choice. The material response is assumed to be characterised by the von Mises plasticity, with a yield stress of $\sigma_y = 243$ MPa. Accordingly, the total strains are additively decomposed into elastic and plastic parts, $\varepsilon_{ij}=\varepsilon_{ij}^e+\varepsilon_{ij}^p$, and the Cauchy stresses are given by, 
\begin{equation}
    \sigma_{ij}=C^e_{ijkl}\varepsilon_{kl}^e
\end{equation}
where $C^e_{ijkl}$ is the elastic isotropic stiffness tensor. Linear isotropic hardening is assumed, with a tangent modulus of $h=2171$ MPa. Accordingly, the yield condition is given by,
\begin{equation}
    f\left( \sigma_{ij} \right) = \sqrt{\frac{3}{2} s_{ij} s_{ij}} - \left( \sigma_y + h \varepsilon_p \right)
\end{equation}
\noindent where $s_{ij}$ is the deviatoric part of the Cauchy stress, $s_{ij}=\mathrm{dev} \left(\sigma_{ij}\right)_{ij}$, and $\varepsilon_p$ is the equivalent plastic strain, $\varepsilon_p = \int \sqrt{2/3 \ \dot{\varepsilon}^p_{ij} \dot{\varepsilon}^p_{ij}} \ \mathrm{d}t $. The rate of the latter evolves as,
\begin{equation}
    \dot{\varepsilon}^p_{ij}=\dot{\lambda} \frac{3}{2\sqrt{3 \ s_{pq} s_{pq}}} s_{ij}
\end{equation}
\noindent with $\dot{\lambda}$ being the plastic multiplier. The loading path is divided into 22 constant increments (using the \textit{Stationary Solver} in COMSOL). The material constants are introduced as properties in the COMSOL \textit{External Material} library and the plastic strain tensor is considered a state variable, initialised with zeros. Finite element calculations are run in both Abaqus and COMSOL using the same user material (\emph{UMAT}) subroutine, with the COMSOL job exploiting the UMAT4COMSOL wrapper. The results obtained are given in Figs. \ref{fig:evpfield} and \ref{fig:evpssc}.\\ 

The equivalent plastic strain contours obtained with COMSOL and Abaqus appear to be indistinguishable, see Fig. \ref{fig:evpfield}. The displacement solution difference between the two approaches is less than $0.6\%$ in terms of relative L2-norm, indicating that the results obtained with Abaqus and COMSOL are identical. The perfect agreement obtained is also evident Fig. \ref{fig:evpssc}a, which shows the engineering stress-strain curves obtained with both packages.

The convergence rates are given in Fig. \ref{fig:evpssc}b. Specifically, the magnitudes of the residuals are provided for each iteration as a function of the load/time increment. As shown in Fig. \ref{fig:evpssc}b, after approximately ten load increments, two iterations are needed to achieve convergence in every increment. This is the same for both Abaqus and COMSOL, and the number of iterations is similar, demonstrating that both provide a similar convergence rate. Some differences are seen in the magnitude of the residuals but this should be considered with care as COMSOL uses the L2-norm of the residual vector to evaluate the convergence while Abaqus uses the $\infty$-norm of the out-of-balance forces.

\subsection{Finite strain neo-Hookean hyperelastic model}
\label{Sec:Hyperelastic}

The case study aims to verify the large strain implementation and showcase the use of UMAT4COMSOL with another material model: non-linear hyperelasticity. To this end, the behaviour resulting from twisting a three-dimensional cube made of a neo-Hookean material is studied. The characteristic length of the cube is 1 m, as in the hyperelasticity example presented in Ref. \cite{Fenics2012}. The constitutive equation of the neo-Hookean hyperelastic model is given by,
\begin{equation}
    \sigma_{i j}=\frac{1}{J}\frac{E}{ 2 \left( 1+\nu \right)} \left(F_{iq}F_{jq}-\frac{1}{3} \delta_{ij} F_{kq}F_{kq} \right)+\frac{E}{3 \left( 1-2\nu \right)}\left( J -1\right) \delta_{ij}
\end{equation}
\noindent and the material parameters used are $E=10^6$ Pa and $\nu=0.3$. These are introduced as properties to the COMSOL \textit{External Material} library. The unitary cuboid is twisted by 60 degrees, and the boundary conditions are given, on one face, by Dirichlet boundary conditions of rotation with respect to the centre and, on the opposite face, by symmetry boundary conditions. For consistency, the calculations are obtained using first-order Lagrange tetrahedral elements with reduced integration in both COMSOL and Abaqus. This element type is denoted C3D4 in Abaqus. Two loading cases are studied: one with a single load increment and another one splitting the load into 10 constant load increments. Calculations are conducted using stationary solvers in both Abaqus and COMSOL. 

\begin{figure}[H]\centering
\includegraphics[width=1\textwidth]{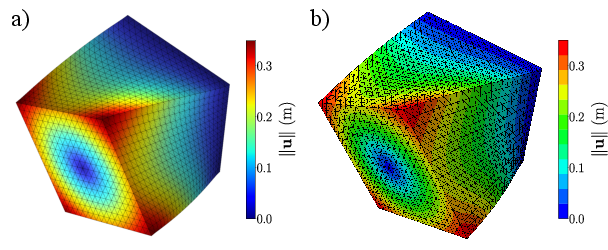}
\caption{Mechanical response of a twisted neo-Hookean cube: contour plots of the displacement field magnitude (L2 norm). Results obtained: (a) with Abaqus, using a non-linear hyperelastic \emph{UMAT}, and (b) with COMSOL, using the same \emph{UMAT} and UMAT4COMSOL.}
\label{fig:neofield}
\end{figure}

The predicted deformed shapes of the twisted cube are given in Fig. \ref{fig:neofield}, showing the contours of the displacement field magnitude. Again, no differences are observed in the results. The contours are also insensitive to the number of load increments employed. Importantly, the relative error in the displacement solution between the Abaqus- and COMSOL-based calculations is below 0.3\%, as measured by the L2-norm. Moreover, this difference does not increase when using only one pseudo-time increment. 

\begin{figure}[H]\centering
\includegraphics[width=0.95\textwidth]{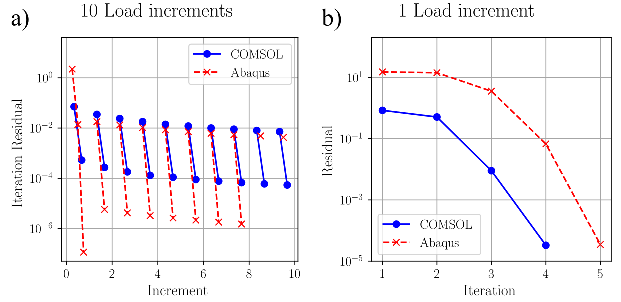}
\caption{Convergence plots of twisting cuboid simulation using a neo-Hookean \textit{UMAT} material in both solvers, Abaqus and COMSOL. Subfigure (a) shows the residual evolution in 10 pseudo-time increments, while subfigure (b) provides convergence plots for 1 pseudo-time increment. In (a), iterations are represented with equispaced subdivisions within each increment interval.}
\label{fig:neocc}
\end{figure}

The non-linear convergence of both solvers is comparable, as shown in Fig. \ref{fig:neocc}, where the magnitude of the residual for each iteration is provided for both the 10-increment (Fig. \ref{fig:neocc}a) and the 1-increment (Fig. \ref{fig:neocc}b) studies. In both loading cases, COMSOL and Abaqus require a similar number of solver iterations to obtain a converged solution. As in the previous case study (Section \ref{Sec:Elastoplasticity}), some quantitative differences can be observed in the magnitude of the residuals, which are intrinsically related to their definition. 

\subsection{Finite strain crystal plasticity}
\label{Sec:CrystalPlasticity}

The last benchmark deals with the deformation of a metallic polycrystal where the material response is characterised using a crystal plasticity material model \cite{Segurado2018}. This benchmark has been chosen for mainly two reasons: (i) its complexity, as the crystal plasticity \textit{UMAT} involves a large number of state variables and exhibits dependencies on the loading path and the strain rate, and (ii) its relevance, as it is not currently possible to run crystal plasticity studies in COMSOL despite its importance in many multi-physics problems. The chosen crystal plasticity material model is elastic-viscoplastic and assumes that plastic deformation only occurs by plastic shear along the slip systems. The deformation gradient is multiplicatively decomposed into elastic and plastic parts,
\begin{equation}
    F_{ij}=F_{ip}^eF_{pj}^p
\end{equation}
For a slip system $\alpha$, the plastic slip rate ($\dot{\gamma}^\alpha$) is given in terms of the resolved shear stress ($\tau^\alpha$) through the following phenomenological power-law,
\begin{equation}\label{eq:gamma_dot}
\dot{\gamma}^\alpha=\dot{\gamma}_0 {\left| \frac{\tau^\alpha}{\tau_c^\alpha} \right|}^n
\end{equation}
where $\dot{\gamma}$ is the reference strain rate, and ${n}$ is an exponent describing the dependence of strain rate on the stress. In Eq. (\ref{eq:gamma_dot}), the critical resolved shear stress on the ${\alpha}$ slip system is the denominator $\tau_c^\alpha$, which evolves following the Asaro-Needleman hardening rule \citep{Peirce1982}, defined by 
\begin{equation}
    \dot{\tau}_c^\alpha=\sum_\beta q_{\alpha\beta} h_0 \sec^2 \left| \frac{h_0\sum_{\delta}\int_0^t\left| \dot{\gamma}^{\delta} \right|\text{d}t}{\tau_s-\tau_0} \right| \dot{\gamma}^\beta
\end{equation}

Accordingly, the plastic part evolves following the sum of plastic slips as:
\begin{equation}\label{eq:fpdot}
\dot{F}^{p}_{ij} = \sum_\alpha \dot{\gamma}^\alpha s^\alpha_i n^\alpha_q \ F^{p}_{qj}
\end{equation}
where $s^\alpha$ and $n^\alpha$ are unit vectors, respectively parallel to the slip direction and normal to the slip plane. This crystal plasticity model represents a face-centred cubic metal lattice with 12 octahedral slip systems and the parameters used are shown in Table \ref{tab:cp}.

\begin{table}[ht!]\centering
\caption{Crystal plasticity parameters adopted in the simulations.}\label{tab:cp}
\resizebox{\textwidth}{!}{\begin{tabular}{lccccccccc}
\hline
Parameter & $C_{11}$ [GPa] & $C_{12}$ [GPa] & $C_{44}$ [GPa] & $h_{0}$ [MPa]  & $\tau_{s}$ [MPa] & $\tau_{0}$ [MPa] & $\dot{\gamma}_{0}$ [-] & $n$ [-] & $q_{\alpha\beta}$ [-] \\
\hline
Magnitude & 168.4  & 121.4 & 75.4  & 541.4 & 109.5 & 60.8 & 0.001 & 10 & 1 \\ 
\hline
\end{tabular}}
\end{table}

\begin{figure}[H]\centering
\includegraphics[width=0.44\textwidth]{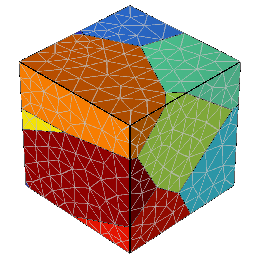}
\caption{Boundary value problem: crystal plasticity-based polycrystal containing 10 grains.}
\label{fig:pxrve}
\end{figure}

\begin{figure}[H]\centering
\includegraphics[width=0.94\textwidth]{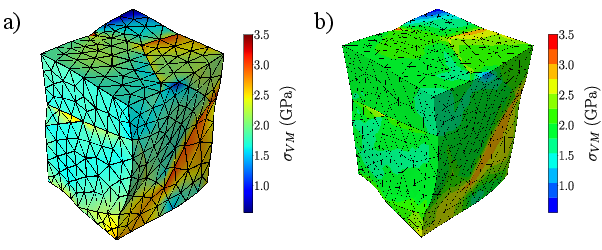}
\caption{Crystal plasticity-based predictions of polycrystalline deformation: contour plots of the von Mises stress (in GPa). Results obtained: (a) with Abaqus, using a crystal plasticity \emph{UMAT}, and (b) with COMSOL, using the same \emph{UMAT} and UMAT4COMSOL. The deformation has been scaled by a factor of 5.}
\label{fig:pxfield}
\end{figure}

\begin{figure}[H]\centering
\includegraphics[width=0.95\textwidth]{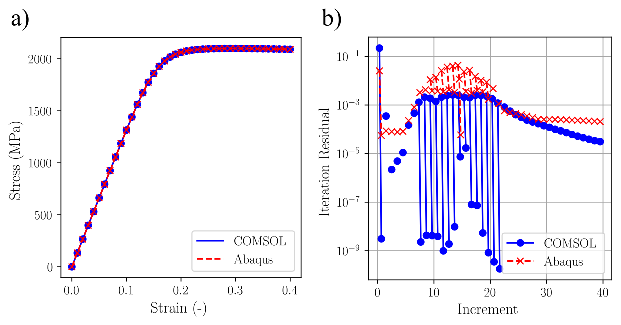}
\caption{Crystal plasticity-based predictions of polycrystalline deformation: (a) resultant macroscopic stress-strain curve, as predicted with Abaqus (+\emph{UMAT}) and with COMSOL (+\emph{UMAT} and UMAT4COMSOL); and (b) convergence plots for both COMSOL and Abaqus solvers, showing the magnitude of the residual for each iteration as a function of the load increment. In (b), iterations are represented with equispaced subdivisions within each increment interval.}
\label{fig:pxssc}
\end{figure}

The boundary value problem is a unit cube containing 10 randomly oriented grains, each behaving as a single crystal with elastic-viscoplastic behaviour, see Fig. \ref{fig:pxrve}. This is a common geometry in the crystal plasticity community \cite{Segurado2018}. One \textit{External Material} must be defined in COMSOL for each grain, as while they all use the same material model, they require different input properties due to the different orientations. Second-order Lagrange tetrahedral elements are used for the polycrystal domain discretisation in both COMSOL and Abaqus (C3D10). Here, it should be noted that to ensure consistency in the number of integration points used, the reduced integration flag must be activated in COMSOL as this implies the use of 4 integration points per element, what Abaqus denotes as a fully integrated quadratic tetrahedral element. The boundary conditions are chosen to mimic a uniaxial strain-controlled tensile test, with the displacement imposed on one face of the cube and the opposite face fixed. A final elongation of $5$\% is imposed. Given the rate-dependent nature of the problem, COMSOL's \emph{Time Dependent} solver is used, with the time being discretised into 20 uniform increments of 1 s. The results obtained are shown in Figs. \ref{fig:pxfield} and \ref{fig:pxssc}. 

As in the previous case studies, the results obtained appear to be identical, independently of the finite element package used. No visible differences are observed in the deformed shape or the stress contours, see Fig. \ref{fig:pxfield}. Due to the different grain orientations, various locations of stress concentration can be observed, and these are present in both COMSOL and Abaqus predictions. From a quantitative viewpoint, the differences in the displacement solution fields observed are below $0.1$\%, validating our approach and implementation. 

The predicted macroscopic stress-strain curve is shown in Fig. \ref{fig:pxssc}a. The results from Abaqus and COMSOL perfectly overlap. The convergence of the simulations is evaluated in Fig. \ref{fig:pxssc}b, which displays the evolution of the equilibrium error estimator. A quadratic or quasi-quadratic rate of convergence is observed. Both COMSOL and Abaqus exhibit a similar performance although on this occasion Abaqus appears to provide a better performance in the last increments. This is likely to be related to the different methods employed for estimating the relative residual (L2-norm in COMSOL and $\infty$-norm in Abaqus).

\subsection{Coupled hydrogen diffusion and deformation in a single crystal}
\label{Sec:Coupled}

Finally, UMAT4COMSOL is used to model a problem of growing scientific and technological interest: the coupling between hydrogen diffusion and material deformation in a single crystal. Hydrogen is famed for `embrittlement' metallic materials, significantly reducing their ductility, fracture toughness and fatigue crack growth resistance, and understanding these hydrogen-material interactions is critical to the safe deployment of hydrogen energy infrastructure \cite{Gangloff2012,EFM2017}. To date, no commercial finite element package enables the coupling of hydrogen transport and crystal plasticity, a key element to gaining fundamental understanding and designing hydrogen-resistant microstructures. As discussed above, COMSOL does not include any crystal plasticity constitutive model, and Abaqus does not have the ability to conduct coupled deformation-diffusion. Thus, the ability of UMAT4COMSOL to break new ground is here showcased by exploiting COMSOL's abilities to run coupled analyses and the existing crystal plasticity material libraries available as Abaqus \emph{UMAT} subroutines.\\

\begin{figure}[H]\centering
\includegraphics[width=0.95\textwidth]{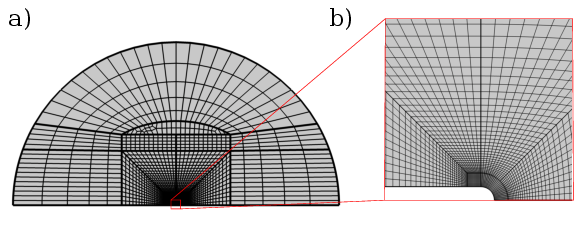}
\caption{Boundary layer formulation and finite element mesh used to describe the coupled deformation-diffusion crack tip behaviour of a single crystal exposed to a hydrogen-containing environment; (a) full geometry, and (b) detail of the crack tip region.}
\label{fig:hcp}
\end{figure}

The analysis focuses on the behaviour of a single crystal containing a blunted crack of opening 0.01 mm. As depicted in Fig. \ref{fig:hcp}, small scale yielding conditions are assumed and a so-called boundary layer formulation is adopted, whereby a remote $K_I$ field is imposed by prescribing the displacement of the outer nodes based on William's linear elasticity solution (see, e.g., Ref. \cite{IJSS2015}). Finite strain conditions are assumed. The outer radius is taken to be sufficiently large so that it has no effect on the solution ($R=150$ mm) and only half of the problem is simulated, taking advantage of symmetry. As shown in Fig. \ref{fig:hcp}, the mesh is refined near the crack tip, to resolve the stress and strain gradients taking place there. This is a classic benchmark in the fracture mechanics community, which has also been comprehensively used in the study of hydrogen embrittlement in isotropic elastic-plastic solids \cite{Sofronis1989,CS2020b}. However, to the best of the author's knowledge, this work constitutes its first application to single crystals. The material behaviour is described by the crystal plasticity model employed in Section \ref{Sec:CrystalPlasticity}, using the same material parameters (Table \ref{tab:cp}). The transport of hydrogen is simulated by using an extended version of Fick's law and assuming a state of equilibrium between the hydrogen located in lattice and trapping sites - so-called Oriani's equilibrium conditions \cite{Oriani1970}. Accordingly, the total hydrogen content $C$ is assumed to additively decomposed into the hydrogen concentration in lattice sites $C_L$ and the hydrogen concentration in trapping sites $C_T$. Only one type of trap is considered: dislocations, and its evolution with the plastic deformation is defined by integrating the crystal plasticity model into the experimentally derived expression first proposed by Kumnick and Johnson \cite{Kumnick1980}, such that the trap density equals,
\begin{equation}
\log N_T =23.26-2.33 \exp{ \left(-5.5\sum_\alpha \gamma^\alpha \right)}
\end{equation}
where $\gamma^\alpha$ is the accumulated plastic shear strain of the crystal plane $\alpha$.\\

Hydrogen diffusion through the crystal lattice is driven by gradients of concentration and of hydrostatic stress $\sigma_h$, bringing a second coupling with the mechanical problem. The balance equation for $C_L$ reads,
\begin{equation} \label{Eq:Hdiffusion}
    \frac{\partial C_L}{\partial t}+\frac{\partial C_T}{\partial t}+\nabla \cdot\left(-D_L \nabla C_L+\frac{D_L \bar{V}_H}{R T} C_L \nabla \sigma_h\right)=0
\end{equation}
where $D_L$ is the lattice diffusion coefficient, $T$ is the absolute temperature, $R$ is the gas constant, and $V_H$ is the partial molar volume of hydrogen. Then, denoting $N_L$ as the density of lattice sites and $W_B$ as the trap binding energy, Oriani's equilibrium results in the following relation between trapped and lattice hydrogen concentrations \cite{hageman2023phase}
\begin{equation}\label{Eq:H_Oriani}
    C_T = N_T \frac{\frac{C_L}{N_L} \exp \left( \frac{W_B}{RT} \right)}{1 + \frac{C_L}{N_L} \exp \left( \frac{W_B}{RT} \right)}
\end{equation}

Equations (\ref{Eq:Hdiffusion}) and (\ref{Eq:H_Oriani}) can be readily considered in COMSOL by either by extending their \textit{Transport of Dilute Species} module (as in Ref. \cite{Diaz2024}), or by using their new (as per version 6.2) hydrogen transport capabilities, based on the implementation by Hageman and Mart{\' i}nez-Pa{\~ n}eda \cite{hageman2022}.\\

\begin{figure}[H]\centering
\includegraphics[width=0.95\textwidth]{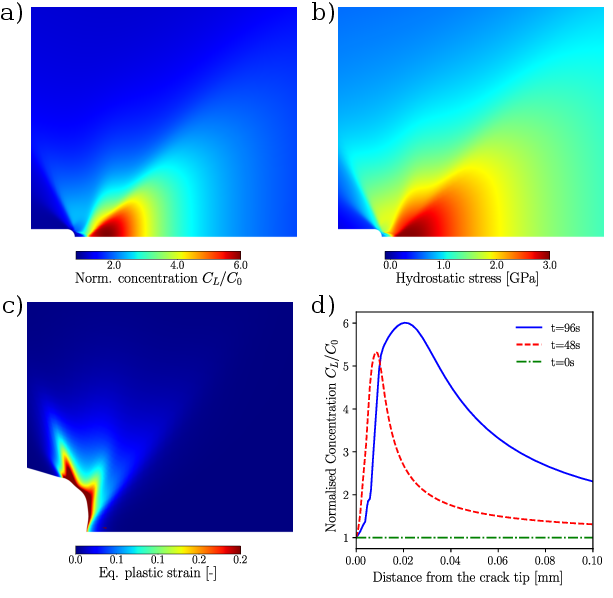}
\caption{Using UMAT4COMSOL to gain insight into the coupled deformation-diffusion behaviour of a single crystal exposed to a hydrogen-containing environment. Near crack tip contours of: (a) normalised lattice concentration $C_L/C_0$, (b) hydrostatic stress $\sigma_h$, and (c) equivalent plastic strain, with the deformation being scaled by a factor of 10. Subfigure (d) shows the quantitative changes in the $C_L$ distribution ahead of the crack tip due to applied load and time.}
\label{fig:hcp2}
\end{figure}

We particularise the model to the study of a single crystal with lattice diffusion coefficient $D_L$=1.27$\times$10$^{-8}$ m$^2$/s, density of lattice sites $N_{L}=8.469$ mol/m$^3$, and partial molar volume $V_H=2\times 10^{-6}$ m$^3$/mol \cite{Sofronis1989}. The trap binding energy associated with dislocations is taken to be $W_B=60$ kJ/mol \cite{Kumnick1980}. The single crystal is oriented parallel to the (001) direction, implying no rotations from the local to the global axis are applied. The temperature is assumed to be $T=300$ K. The outer displacement is linearly increased with time until reaching an applied intensity factor of $K_I=65.8$ MPa$\sqrt{\text{m}}$ at 96 s. The hydrogen concentration initial and boundary conditions are as follows; no flux is assumed everywhere except for the crack surface, where a fixed concentration of $C_0=0.00346$ mol/m$^3$ is imposed, which is also the initial hydrogen concentration in the metal lattice. The computational domain is discretised with second-order Lagrange triangular elements with full integration and a multi-pass staggered (segregated) solution scheme is adopted. No comparison with Abaqus is provided, as this sort of analysis cannot be conducted using only \emph{UMAT} subroutines (a UMATHT \cite{AM2020} or UEL \cite{TAFM2020c} are needed). The results obtained are shown in Fig. \ref{fig:hcp2}, in terms of contours of normalised hydrogen concentration ($C_L/C_0$), hydrostatic stress ($\sigma_h$) and equivalent plastic strain $\varepsilon_p=\sum_\alpha \gamma^\alpha$.  \\ 

The results obtained, given in Fig. \ref{fig:hcp2}, show the expected trends. The lattice hydrogen content (Fig. \ref{fig:hcp2}a) is highest in the regions of high hydrostatic stress (Fig. \ref{fig:hcp2}b). Since a conventional model is used, the stress and concentration peaks are slightly ahead of the crack tip; this would change if the role of strain gradients is accounted for \cite{niordson2014computational,IJHE2016}. The equivalent plastic strain is highest near the crack tip, as shown in Fig. \ref{fig:hcp2}c, where the deformation has been scaled by a factor of 10. Accordingly, the hydrogen content trapped in dislocations (not shown) also increases as the crack tip is approached. Finally, Fig. \ref{fig:hcp2}d shows more quantitative estimates of normalised lattice hydrogen content ahead of the crack tip, showing the expected sensitivity of $C_L$ to time and to the applied mechanical load. In terms of the convergence, the rate obtained is similar to those when using built-in elasto-plastic COMSOL materials.

\color{black}

\section{Conclusions}
\label{Sec:Conclusions}

We have presented UMAT4COMSOL, a new wrapper that links COMSOL C-coded \emph{External Material} library with Abaqus Fortran-coded user material (\textit{UMAT}) subroutines. This enables bringing to the multi-physics environment of COMSOL the advanced material models that the solid mechanics community has developed and implemented through Abaqus \textit{UMAT} subroutines. Thus, UMAT4COMSOL enables the combination of advanced multi-physics and material modelling tools, as necessary to predict the degradation of Li-Ion batteries, the corrosion of metals and the oxidation of superalloys, among others. We validate our framework and demonstrate its potential by addressing three case studies of particular relevance: (i) the elastoplastic behaviour of a holed plate, (ii) the finite strain non-linear hyperelastic deformation of a twisted cube, and (iii) the crystal plasticity-based prediction of microscopic and macroscopic deformation in a polycrystalline solid. The results show that predictions obtained using COMSOL and UMAT4COMSOL are identical to those obtained using Abaqus and the associated \emph{UMAT} subroutine. Moreover, convergence rates also exhibit a remarkable agreement, despite the different characteristics of each solver. To further highlight the potential of UMAT4COMSOL, the coupled deformation-diffusion behaviour of a single crystal exposed to a hydrogen-containing environment is simulated using a crystal plasticity \emph{UMAT} subroutine and COMSOL's multi-physics capabilities. This analysis, of increasing technological importance, highlights the ability of the wrapper presented to assist in extending the state-of-the-art in materials and coupled physical modelling.

\section*{Acknowledgements}

The authors gratefully acknowledge financial support through grant EP/V038079/1 (``SINDRI") from the Engineering and Physical Sciences Research Council (EPSRC). Sergio Lucarini acknowledges financial support from the Marie Skłodowska-Curie Individual European Fellowship under the European Union's Horizon 2020 Framework Programme for Research and Innovation through the project SIMCOFAT (Grant agreement ID: 101031287). Emilio Mart{\' i}nez-Pa{\~ n}eda acknowledges financial support from UKRI’s Future Leaders Fellowship programme [grant MR/V024124/1].

%\appendix

\section*{Data availability}
The software developed, UMAT4COMSOL, is made freely available together with example case studies and documentation. UMAT4COMSOL can be downloaded from the website of the research group (\url{https://www.imperial.ac.uk/mechanics-materials/codes/}), a website repository (\url{www.empaneda.com/codes}) and a GitHub repository (\url{https://github.com/sergiolucarini}).

\end{document}